\documentclass{aastex62}

\received{}
\revised{}
\accepted{2018/06/27}
\submitjournal{ApJ}

\shorttitle{In-situ generation of transverse MHD waves from colliding flows in the solar corona}
\shortauthors{Antolin et al.}

\begin{document}

\title{In-situ generation of transverse MHD waves from colliding flows in the solar corona}

\correspondingauthor{Patrick Antolin}
\email{patrick.antolin@st-andrews.ac.uk}

\author[0000-0003-1529-4681]{Patrick Antolin}

\author[0000-0001-5274-515X]{Paolo Pagano}

\author[0000-0002-1452-9330]{Ineke De Moortel}

\affiliation{School of Mathematics and Statistics, University of St. Andrews, St. Andrews, Fife KY16 9SS, UK}

\author[0000-0001-6423-8286]{Valery M. Nakariakov}

\affiliation{Centre for Fusion, Space and Astrophysics, University of Warwick, Coventry CV4 7AL, UK}
\affiliation{School of Space Research, Kyung Hee University, Yongin, 446-701 Gyeonggi, Korea}



\begin{abstract}
Transverse MHD waves permeate the solar atmosphere and are a candidate for coronal heating. However, the origin of these waves is still unclear. In this work, we analyse coordinated observations from \textit{Hinode}/SOT and \textit{IRIS} of a prominence/coronal rain loop-like structure at the limb of the Sun. Cool and dense downflows and upflows are observed along the structure. A collision between a downward and an upward flow with an estimated energy flux of $10^{7}-10^{8}$~erg~cm$^{-2}$~s$^{-1}$ is observed to generate oscillatory transverse perturbations of the strands with an estimated $\approx40~$km~s$^{-1}$ total amplitude, and a short-lived brightening event with the plasma temperature increasing to at least $10^{5}~$K. We interpret this response as sausage and kink transverse MHD waves based on 2D MHD simulations of plasma flow collision. The lengths, density and velocity differences between the colliding clumps and the strength of the magnetic field are major parameters defining the response to the collision. The presence of asymmetry between the clumps (angle of impact surface and/or offset of flowing axis) is crucial to generate a kink mode. Using the observed values we successfully reproduce the observed transverse perturbations and brightening, and show adiabatic heating to coronal temperatures. The numerical modelling indicates that the plasma $\beta$ in this loop-like structure is confined between $0.09$ and $0.36$. These results suggest that such collisions from counter-streaming flows can be a source of in-situ transverse MHD waves, and that for cool and dense prominence conditions such waves could have significant amplitudes.

\end{abstract}



\section{Introduction} \label{sec:intro}

Transverse MHD waves permeate the solar atmosphere and constitute a possible candidate for coronal heating \citep[for a review, see for example][]{Arregui_2012LRSP....9....2A,DeMoortel_Nakariakov_2012RSPTA.370.3193D,Arregui_2015RSPTA.37340261A}. A main source of evidence of these waves comes from observations of prominences and coronal rain, in which the naturally cold, dense and optically thicker plasma conditions allow much higher spatial resolution and reduced line-of-sight confusion  \citep{Lin_2005SoPh..226..239L,Lin_2011SSRv..158..237L, Okamoto_2007Sci...318.1577O,Ning_2009AA...499..595N,Hillier_2013ApJ...779L..16H,Schmieder_2013ApJ...777..108S,Okamoto_2015ApJ...809...71O,Vial_Engvold_2015ASSL..415.....V}. However, the origin of these waves (in coronal and prominence structures) remains unclear and is usually assumed to be in convective motions, or through mode conversion of $p-$modes propagating from the solar interior.

Another commonly observed feature of cool coronal structures such as prominences and rainy loops are field-aligned flows with speeds of $10-100$~km~s$^{-1}$ and $40-200$~km~s$^{-1}$, respectively \citep{Ofman_Wang_2008AA...482L...9O,Antolin_Rouppe_2012ApJ...745..152A,Alexander_2013ApJ...775L..32A,Kleint_2014ApJ...789L..42K}. Both downflows and upflows are observed along the legs of prominences \citep{Vial_Engvold_2015ASSL..415.....V}. These longitudinal dynamics are commonly associated to the formation mechanism of prominences or coronal rain, such as thermal instability or thermal non-equilibrium
\citep{Antiochos_1999ApJ...512..985A,Karpen_etal_2001ApJ...553L..85K,Antolin_2010ApJ...716..154A,Xia_2017AA...603A..42X}. 

Through coordinated observations of a prominence/coronal rain complex (\S\ref{sec:obs}) with \textit{Hinode} \citep{Kosugi_2007SoPh..243....3K}, the \textit{Interface Region Imaging Spectrograph} \citep[IRIS;][]{DePontieu_2014SoPh..289.2733D} and the \textit{Solar Dynamics Observatory}\citep[SDO;][]{Pesnell_2012SoPh..275....3P}, and numerical MHD modelling (\S\ref{sec:model}), we show that in-situ collisions from such counter-streaming flows could be a source of transverse MHD waves in the corona.

\section{Observations}\label{sec:obs}

\begin{figure}
\centering
\includegraphics[scale=0.5,bb=0 0 572 556]{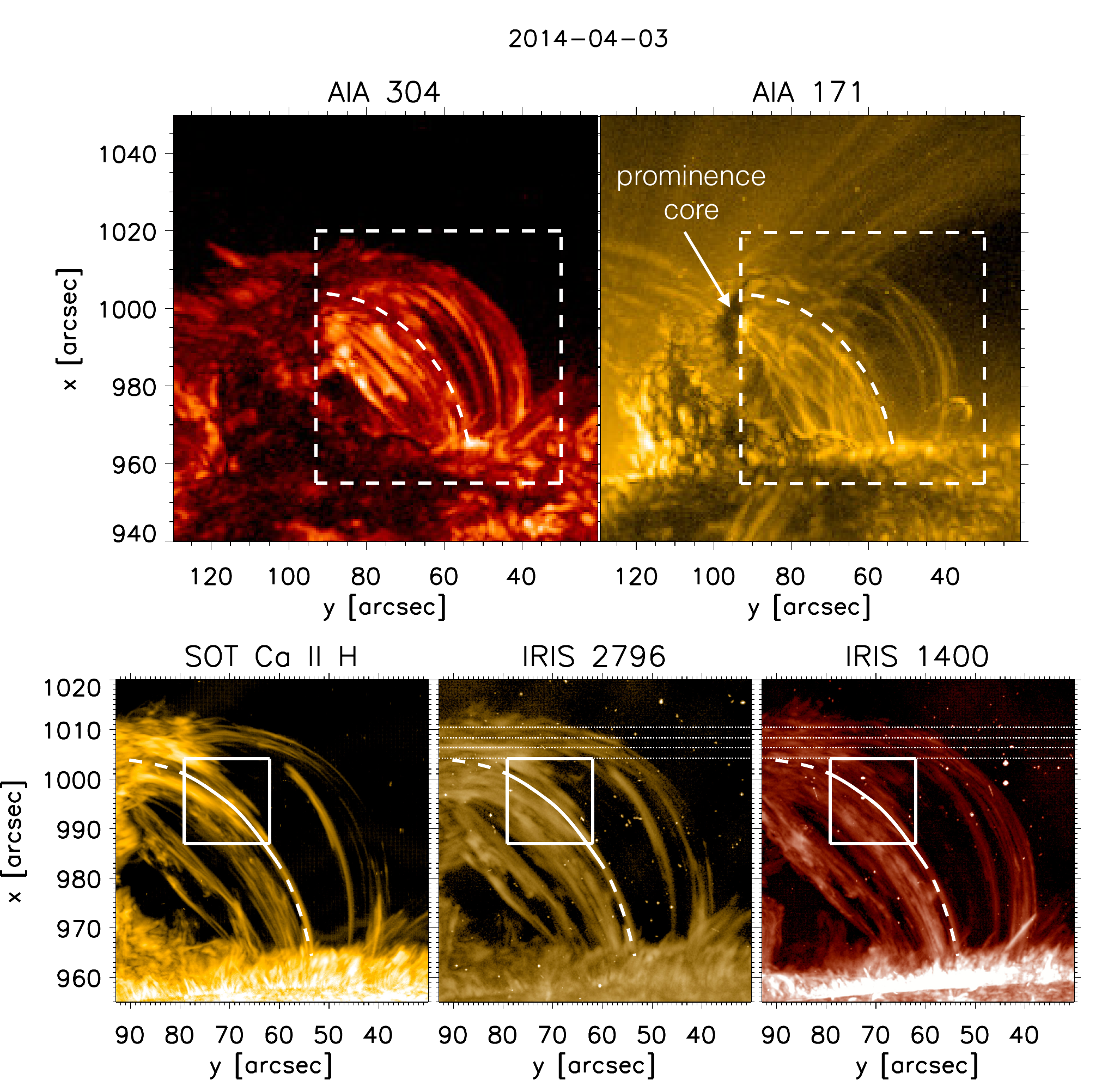}
\caption{%Prominence/coronal rain complex at the West limb of the Sun co-observed with \textit{Hinode}/SOT, \textit{IRIS} and \textit{SDO}/AIA.
From top left to bottom right we show variance images (sum of squared differences from a 8~min average image) in the AIA 304, AIA 171, in the SOT \ion{Ca}{2}~H, SJI 2796 and SJI 1400 of \textit{IRIS}.
%The AIA images show the entire prominence/coronal rain complex.
The FOV of the bottom three panels is shown as a dashed square in the top 2 panels.
%A high standing prominence core can be distinguished in EUV absorption in AIA 171, whose edge is visible in the top left in each of the bottom panels. This prominence core is connected to the solar surface through loop-like structures.
A particular set of downflows is followed along one of these structures (white dashed/solid curve).
%An enlarged image of the white solid square is shown in Fig.~\ref{obs2}.
The horizontal dotted lines in the IRIS panels indicate the location of the slit during the 4-step raster. Note that the solar East-West direction corresponds to the vertical axes in the panels.}
\label{obs1}
\end{figure}

On April 3rd, 2014, \textit{SDO}, \textit{IRIS} and \textit{Hinode} co-observed a prominence/coronal rain complex on the West limb of the Sun (\textit{IRIS-Hinode} operation plan 254), shown in Fig.~\ref{obs1}. The \textit{Hinode} Solar Optical Telescope \citep[SOT;][]{Suematsu_2008SoPh..249..197S, Tsuneta_2008SoPh..249..167T} observed from 13:16UT to 14:30UT in the \ion{Ca}{2}~H line, with a cadence of 8~s (1.23~s exposure), with $0.^{\prime\prime}109$~pixel$^{-1}$ platescale, and a field-of-view (FOV) of $111\arcsec\times111\arcsec$, centred at helioprojective coordinates $(x,y)=(996.5,31.1)$. \textit{IRIS} observed from 13:16UT to 14:53UT with a 4-step sparse raster program (OBS ID 3840259471), with a cadence for the Slit-Jaw Imager (SJI) of 18.27~s (exposure time of 8~s) and 9~s for the spectrograph SG (roughly 37~s per raster position), with $0.^{\prime\prime}166$~pixel$^{-1}$ platescale, a FOV of $127\arcsec\times128\arcsec$ centred at $(x,y)=(1007.1,34.2)$, containing the SOT FOV.
%\textit{IRIS} was not rolled, leading to a slit parallel to the solar limb and about $55\arcsec$ above it.
The \textit{IRIS} observing program included both the SJI 2796 and SJI 1400 filtergrams, which are dominated by \ion{Mg}{2}~k emission at 2796.35~\AA~around $2\times10^{4}$~K and \ion{Si}{4} emission at 1402.77~\AA~around $10^{5}~$K, respectively. The images from the Atmospheric Imaging Assembly \citep[AIA;][]{Lemen_etal_2011SoPh..tmp..172L} were in level 1.5. Their cadence is $12~$s, with a platescale of $0.^{\prime\prime}6$~pixel$^{-1}$.

The SOT dataset was processed using the $\tt{FG\_PREP}$ Solarsoft routine. The SJI data corresponds to level~2 data \citep{DePontieu_2014SoPh..289.2733D}, which includes correction for thermal variations of the pointing by co-aligning each image using a cross correlation maximisation routine. SOT, \textit{IRIS} and AIA data were co-aligned manually.
%using mostly the on-disk solar features. For SOT and IRIS we used, respectively, \ion{Ca}{2}~H and \ion{Mg}{2}~k (of the SJI 2796), which form in similar regions of the atmosphere. For IRIS and AIA we used the \ion{Si}{4}~1402.77~\AA\, (of the SJI 1400) and \ion{He}{2} (of the AIA 304, together with EUV absorption features in the AIA 171), which are both transition region lines. 

The target of the observations was a loop-like structure stemming from a high ($\approx55\arcsec$ above the limb) prominence core, reminiscent of the model by \citet{Keppens_2014ApJ...789...22K}.
%Due to its similarity to coronal rain, such structures are also known as prominence/coronal rain complexes \citep{Liu_2016SPD....47.0402L}.
In Fig.~\ref{obs1} we show a variance image of this structure in \ion{Ca}{2}~H, SJI 2796 and SJI 1400, where the variance is taken over the first 8~min of the observation.
%On the left hand side of the image the edge of the prominence core can be seen, and 3 legs apparently connecting it with the solar surface display continuous plasma motions. 

\section{Colliding flows}\label{sec:collision}

\begin{figure}
\centering
\includegraphics[scale=0.5,bb=0 0 530 558]{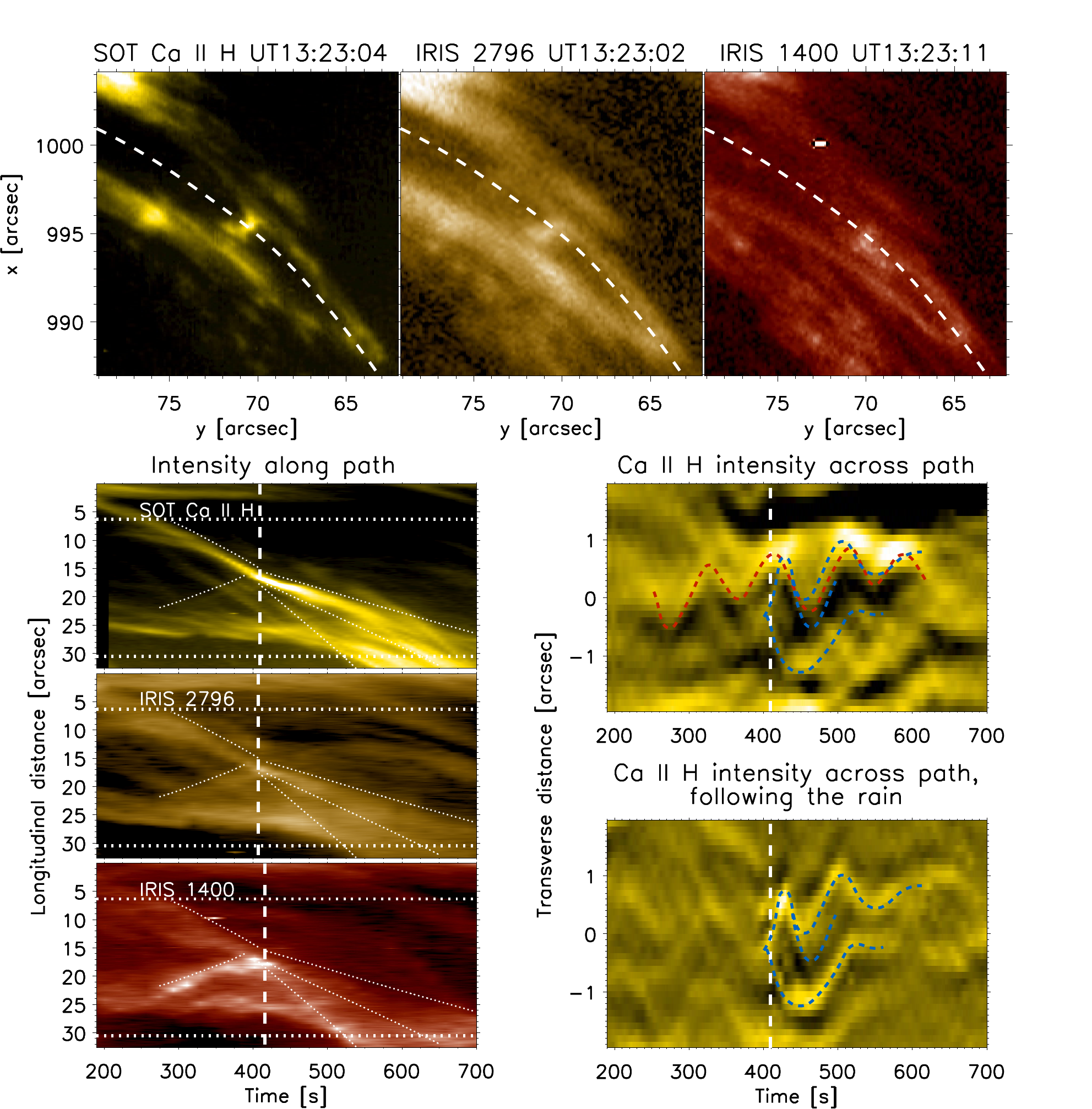}
\caption{%Collision of a downflowing and upflowing set of clumps leading to transverse MHD waves.
The 3 top panels show, from left to right, the region indicated by the small white square (solid lines) of Fig.~\ref{obs1}, in the \ion{Ca}{2}~H line of SOT, the SJI 2796 and SJI 1400 of \textit{IRIS}. The white dashed curve shows the path of counter-streaming plasma clumps. The time-distance diagram along this path is shown in the 3 left panels, where distance is measured from the top of the path shown in Fig.~\ref{obs1}. The intensity is integrated over a width of $1\arcsec$. The horizontal dotted lines in these panels indicate the top and bottom coordinates along the trajectory within the FOV of the top 3 panels. The vertical dashed line indicates the closest time to collision for each channel.
%between a downflow and an upflow with POS speeds around $40-50$~km~s$^{-1}$. The downflowing material's speed is affected, as evidenced by the different slopes of the overlaid white lines.
The middle and bottom right panels show transverse cuts to the trajectory of the clumps.
%evidencing transverse MHD waves.
The contours of the brightest oscillatory paths are traced visually in dashed red and blue curves, respectively for the middle and bottom right panels. In the middle panel the intensity along part of the path (solid curve in Fig.~\ref{obs1}) is integrated for each transverse cut. In the lower panel the downflowing set of clumps is followed and at each time the transverse cut to the trajectory is plotted. See also the accompanying animation.}
\label{obs2}
\end{figure}

The focus of the present investigation is the main loop-like structure connected to the prominence core seen in the middle of the images in Fig.~\ref{obs1}.
The flows along this loop structure are clumpy and multi-stranded, particularly in the higher resolution \ion{Ca}{2}~H intensity images, a general characteristic of coronal rain material when observed at high resolution \citep{Antolin_2015ApJ...806...81A}. Mostly downflows are observed, stemming from the prominence core (top left) towards the surface. Additionally, and in contrast to the usual coronal rain, the current case also exhibits a significant amount of upflows probably caused by dips at the loop apex, enhancing thermal instability. 

We follow a particular set of clumps during their downward trajectory (dashed curve in Figs.~\ref{obs1} and \ref{obs2}) at a constant velocity of $\approx50$~km~s$^{-1}$ in the plane-of-the-sky (POS). At high resolution in \ion{Ca}{2}~H, the clumps are $0.3\arcsec-1.4\arcsec$ in width (across the loop), with a continuously variable length of a few arcsec. Half-way along the loop at $t=13.23$UT the clumps' intensities in all 3 channels suddenly increase, and a bright front of about $2\arcsec$ in width is observed (visible in Fig.~\ref{obs2}). Afterwards, the downward speeds of some of the clumps are reduced by half, as seen in the time-distance diagram along the loop (3-set bottom left panels in Fig.~\ref{obs2}). In addition, the clumps are seen to oscillate transversely in \ion{Ca}{2}~H
following the intensity increase. This is best seen in the time-distance diagram transverse to the loop (bottom right panel in Fig.~\ref{obs2}), where we follow the clump along its downward trajectory. We can see that some clumps undergo an outward transverse motion of $\approx1\arcsec$ (radially away from the Sun, positive transverse distance in the panel), while others undergo an inwards transverse motion of similar amplitude. The initial transverse velocity in the POS is $\approx25$~km~s$^{-1}$. The outward transverse motion can be tracked for longer times and the oscillation is damped in $2-3$ periods. Note that the period of the oscillation increases from the first to the second oscillation, from $50-60~$s to $80-90$~s.  

The \textit{IRIS} SJIs reveal an upward flow with a speed of $\approx40~$km~s$^{-1}$ that seems to collide with the downward flow (see the time-distance diagram along the loop). The time of collision coincides with both the brightness increase in all 3 channels and the start of the transverse oscillation. This upward flow is clearly visible in SJI 1400, but barely visible in SJI 2796 and \ion{Ca}{2}~H. This intensity difference across the channels suggests that the upward flow is about 10 times hotter, with a temperature around $10^{5}~$K. This event suggests that a collision occurs between the downward and upward flows, which then leads to the generation of transverse MHD waves. 

The middle right panel in Fig.~\ref{obs2} shows the presence of transverse MHD waves along the loop even prior to the flow collision, with a period of $\approx90-100$~s. The oscillation initiated by the collision (blue curves in the panels) is different from this background oscillation (red curves). For instance, the time of flow collision ($t\approx410$~s) and the subsequent maxima is initially out-of-phase with this background oscillation. The increasing period of the generated transverse oscillation leads to in-phase second maxima.

We have estimated the total densities of the plasma towards the prominence core with the help of the AIA data. These estimates are based on EUV absorption by the cool material (mainly neutral Hydrogen and Helium) in the AIA wavelengths following the technique by \citet{Landi_2013ApJ...772...71L} and \citet{Antolin_2015ApJ...806...81A}. Taking a 5\% Helium abundance, we find values between $6\times10^{10}$~cm$^{-3}$ and $3\times10^{11}$~cm$^{-3}$, in agreement with previous measurements in prominences \citep{Vial_Engvold_2015ASSL..415.....V} and coronal rain \citep{Antolin_2015ApJ...806...81A}. 

\begin{figure}
\centering
\includegraphics[scale=0.7,bb=0 8 304 490]{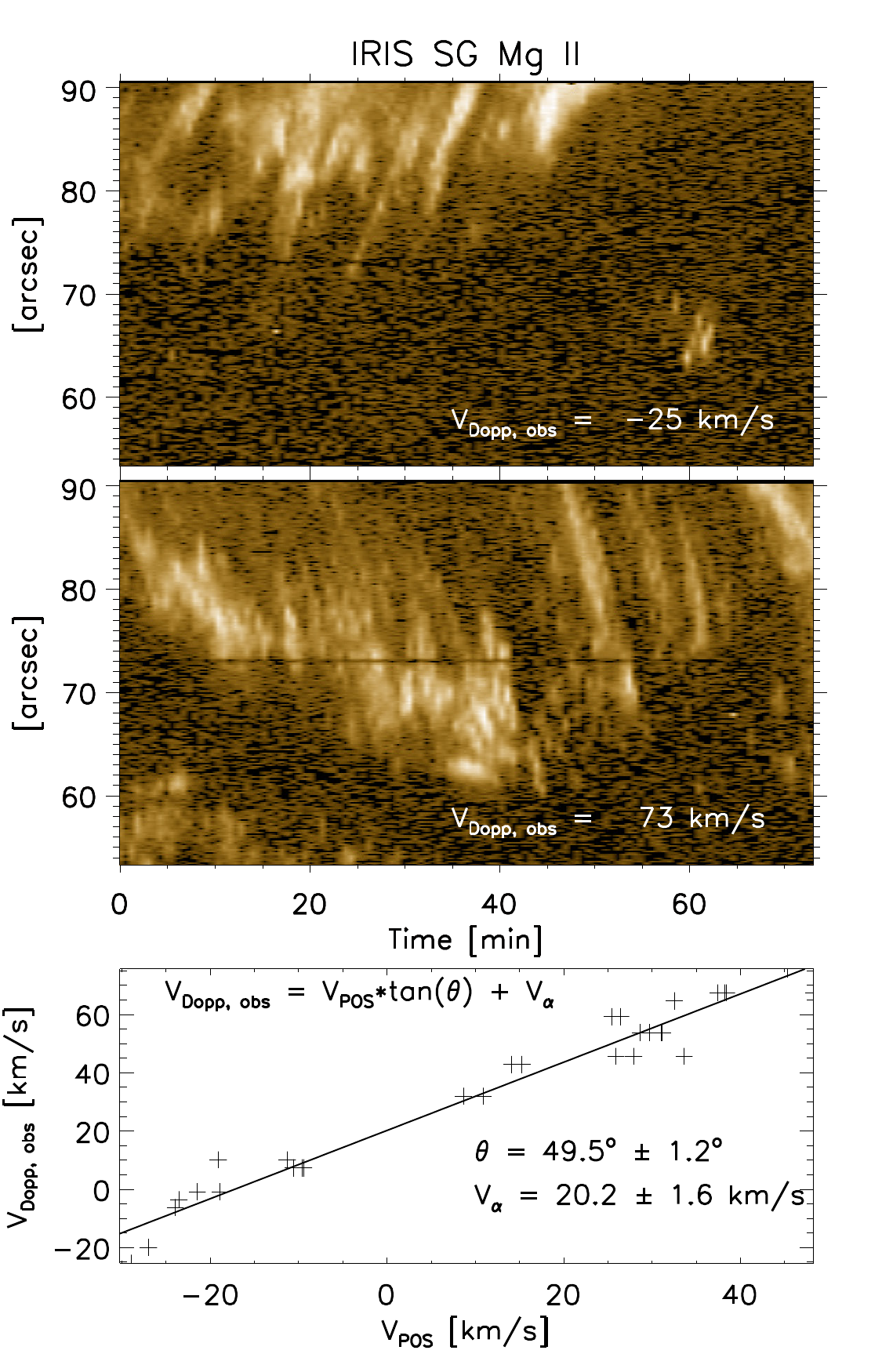}
\caption{The top 2 panels show the time-distance diagram along part of the bottom-most slit position (tangent to the loop apex) at 2 wavelength positions. The slopes ($v_\mathrm{POS}$) of several paths of clumps are measured for each wavelength position ($v_\mathrm{Dopp,obs}$) and plotted in the bottom panel. Positive/negative slopes correspond, respectively, to negative/positive $V_\mathrm{POS}$.}
\label{obs3}
\end{figure}

Due to its position and orientation the IRIS slit captures a significant fraction of the loop near its apex (see Fig.~\ref{obs1}). Hence, the slit captures the spectra of several flows directed along the loop. These flows produce positive / negative slopes for upward / downward flows, respectively, in the time-distance diagram along the slit (see Fig.~\ref{obs3}). For each wavelength position \citep[corresponding to a Doppler velocity $v_\mathrm{Dopp, obs}$, assuming the wavelength value from CHIANTI for the zero velocity,][]{Dere_etal_2009AA...498..915D} we select the most distinct paths and measure the slopes, which corresponds to the POS velocity along the slit $v_\mathrm{POS}$. These measurements are shown in the scatter plot of the bottom panel. The quantities follow the relation $v_\mathrm{Dopp,obs} = v_\mathrm{POS} tan (\theta) + v_\mathrm{\alpha}$, where $v_\mathrm{\alpha}$ corresponds to the zero Doppler velocity in the reference frame of the loop, and $\theta$ corresponds to the angle of the flow path (at the loop apex) with the POS plane. We find $\theta=49.5^{\circ}\pm1.2^{\circ}$ and $v_\mathrm{\alpha}=20.2\pm1.6$~km~s$^{-1}$. This implies that the total velocity of a flow with POS velocity of 50~km~s$^{-1}$ is close to 77~km~s$^{-1}$, and that the transverse amplitude of the wave is $\approx40~$km~s$^{-1}$.

\section{MHD Model}
\label{sec:model}

In order to interpret our observations,
we set up a 2D MHD model of counter-streaming plasma clumps.
%where two clumps of plasma
%flow along the same bundle of field lines in opposite direction.
%The collision/interaction of these plasma clumps leads to the deformation of the magnetic field and the generation of a kink wave long the magnetic field lines.

We consider a 2D spatial domain that extends for 
$12~$Mm in the $x-$direction (field aligned)
and $6~$Mm in the $y-$direction. %(perpendicular to the magnetic field and flows).
The magnetic field, $B_0$, is uniform and directed along the $x-$direction.
Two trapezoidal clumps are placed at a distance of $4~$Mm
and are $1~$Mm wide and $3~$Mm long
in a background corona where the density is $n_0=1.2\times10^9$ $cm^{-3}$
and the temperature is $T_0=1~$MK.
The clumps are $n_\mathrm{c}$ times denser than the background plasma and
colder in order to maintain pressure equilibrium.
The plasma within the clumps has an initial velocity of $V_\mathrm{B}=\pm70~$km~s$^{-1}$.
The shape of the clumps is such that the two facing sides are inclined 
in the same direction with an angle $\phi=50^{\circ}$.
We numerically solve the set of ideal MHD equations using
the MPI-AMRVAC software~\citep{Porth_2014ApJS..214....4P}.
We neglect non-ideal effects, as they would act on time scales longer than the observed evolution.

\begin{figure}
\centering
\includegraphics[scale=0.25,bb=0 0 908 567]{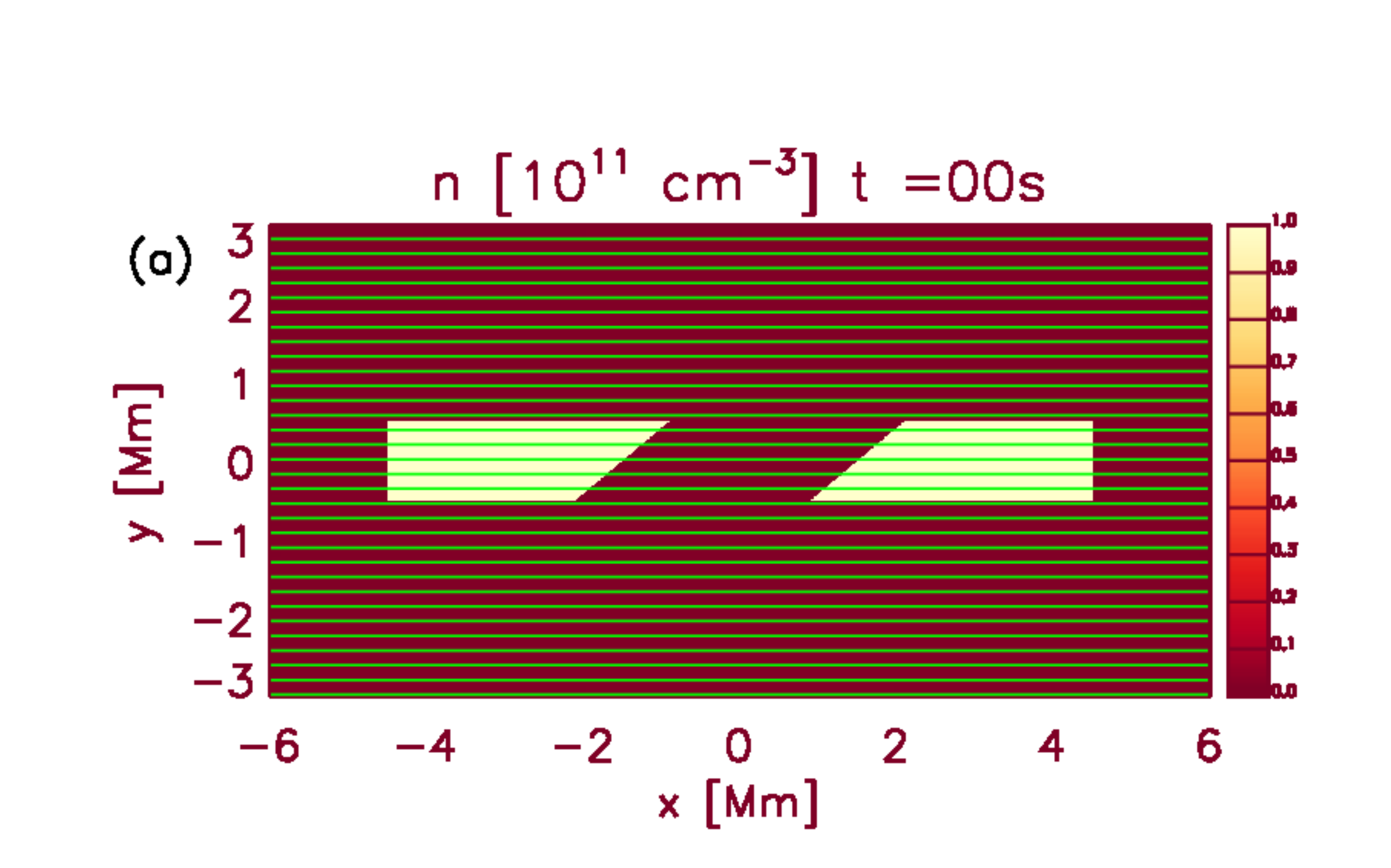}
\includegraphics[scale=0.25,bb=0 0 908 567]{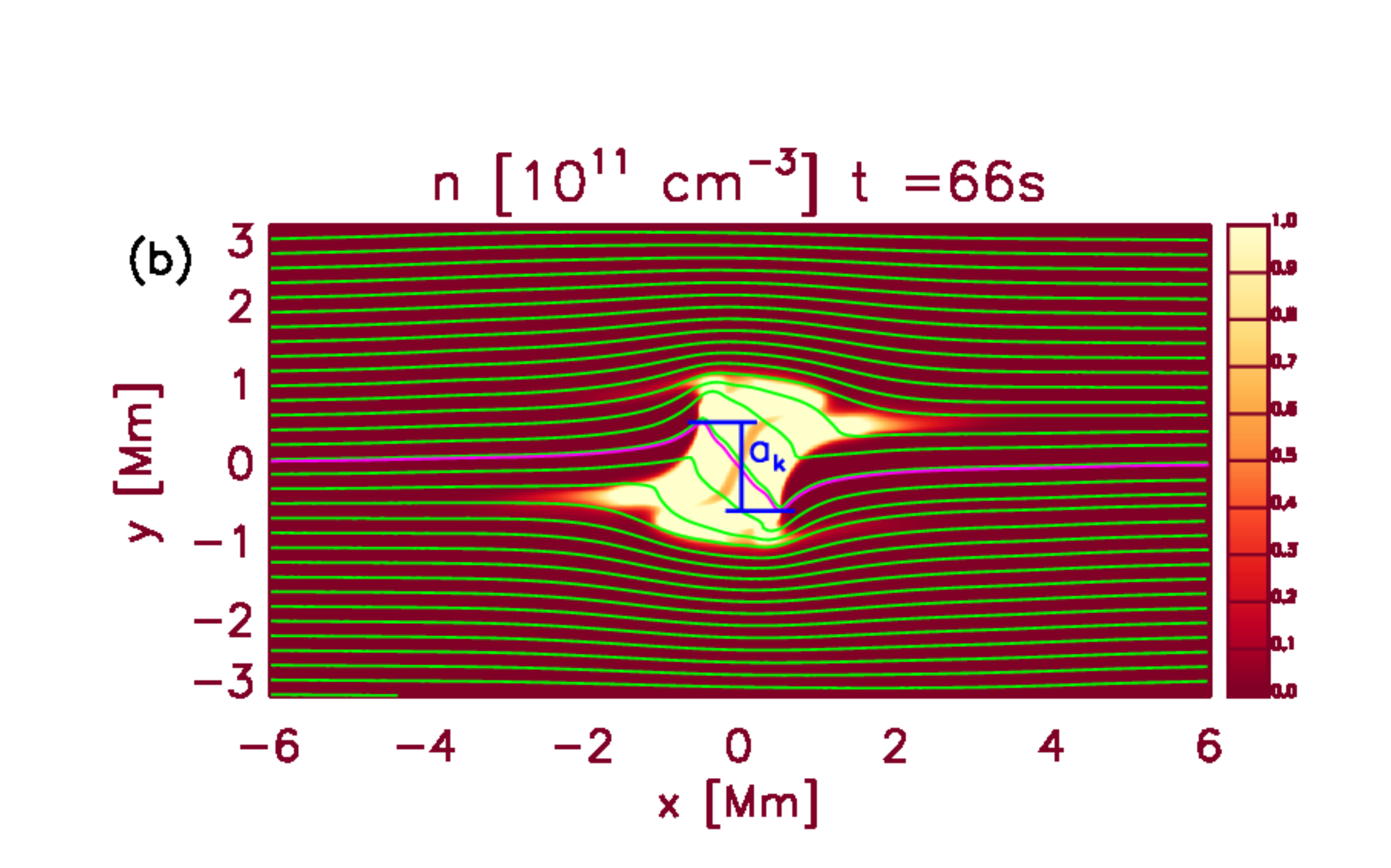}
\includegraphics[scale=0.25,bb=0 0 908 567]{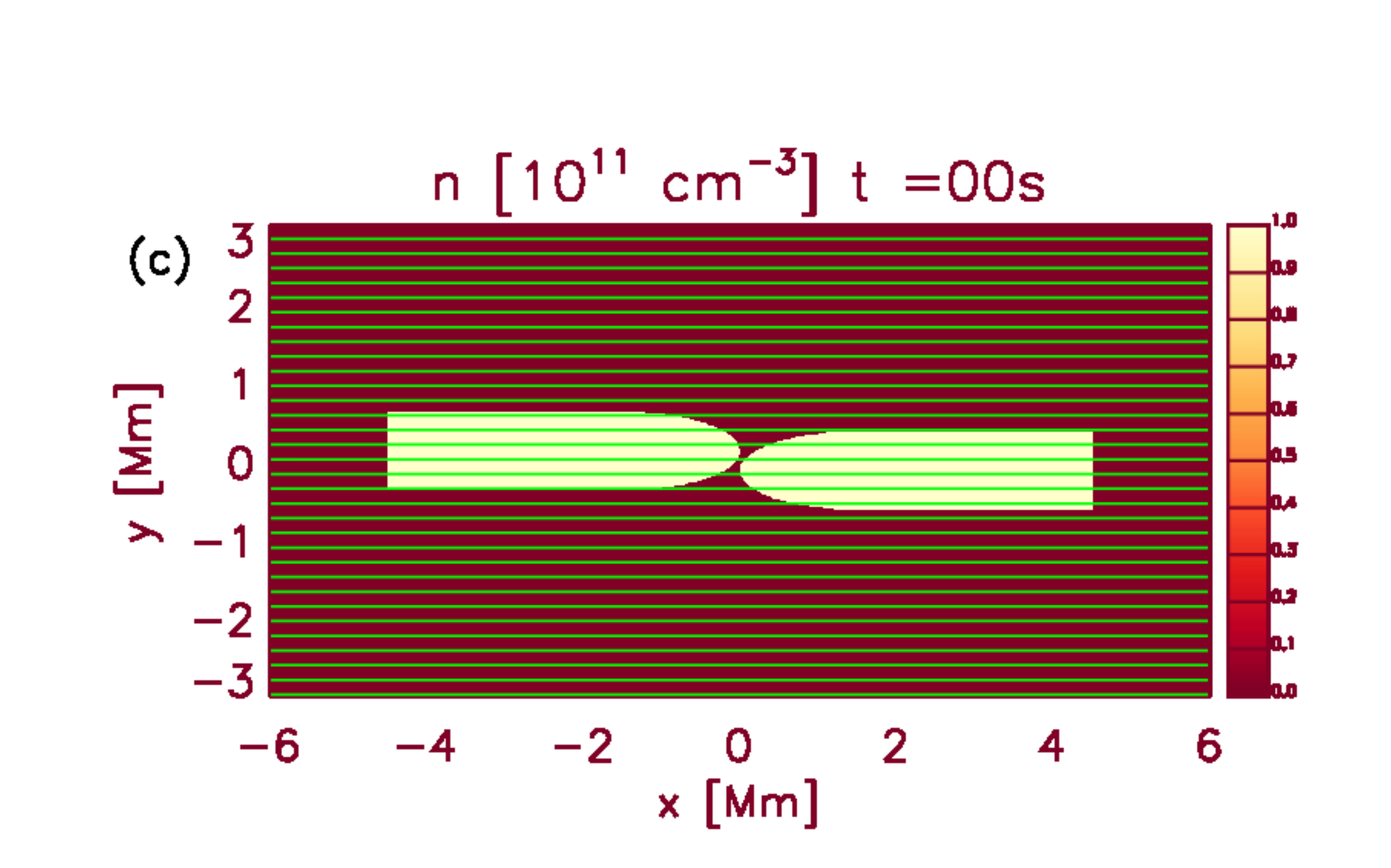}
\includegraphics[scale=0.25,bb=0 0 908 567]{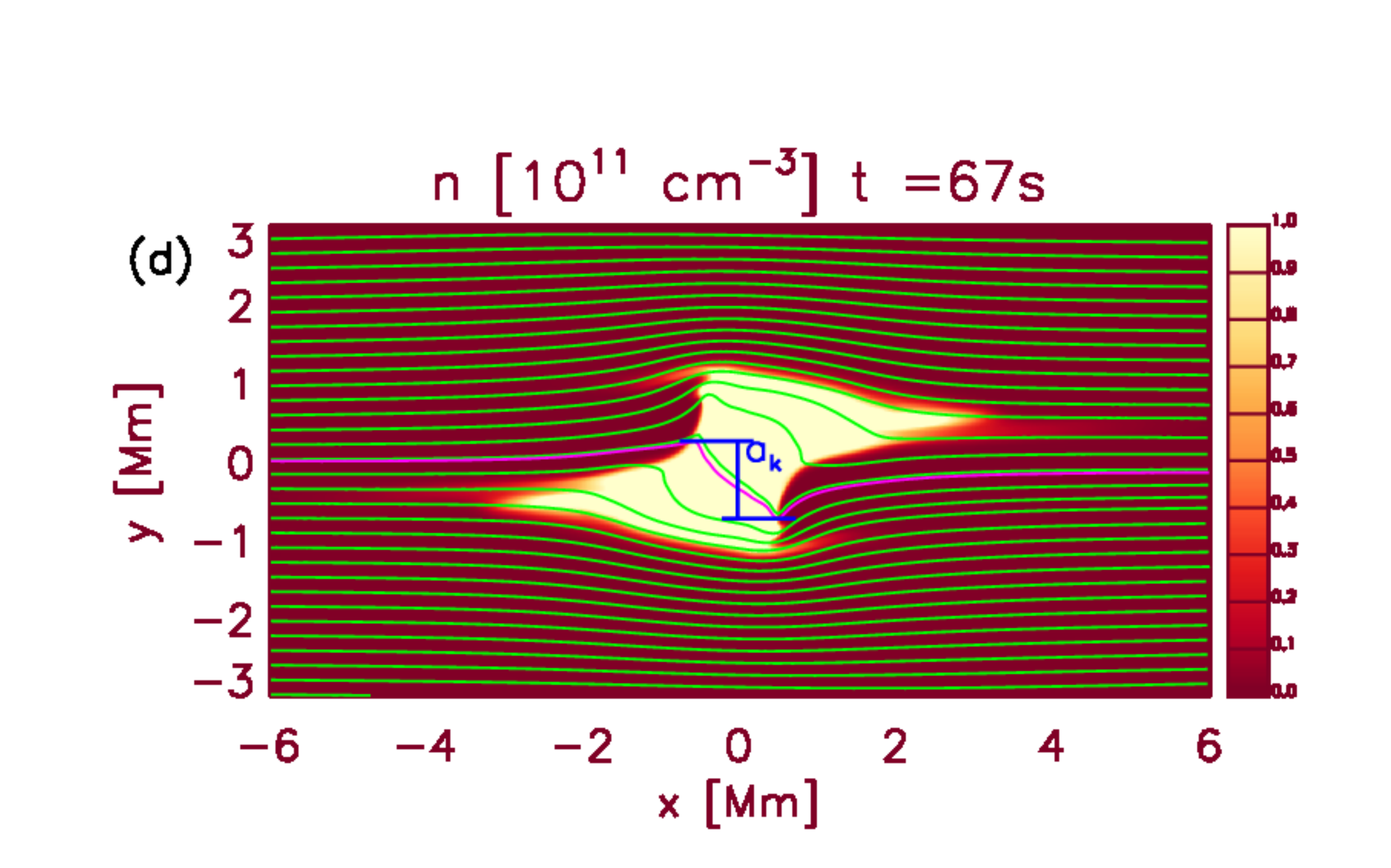}
\caption{Maps of number density with overlaid magnetic field lines for the initial conditions of the simulations with trapezoidal or circular front clumps ((a) and (c)) and the times when the kink is maximum ((b) and (d)), for the strong collision scenario when $\beta=0.098$.
The blue lines mark the amplitude $a_k$ of the kink. See also the accompanying animation.}
\label{simak}
\end{figure}

The observations suggest lower and upper limits for the density contrast of the clumps.
Therefore, we model
a strong collision scenario where we have $n_\mathrm{c}=100$ and
a weak collision scenario where we have $n_\mathrm{c}=25$.
For each scenario we first run 4 simulations with different values of plasma $\beta$
($0.02, 0.05, 0.2, 0.5$)
that defines a posteriori the field strength $B_0$.
Fig.~\ref{simak}a illustrates the density and magnetic field lines 
in the initial condition for all simulations.

The clumps move towards one another and leads to the compression of plasma between the clumps, which is adiabatically heated up to $\sim3$~MK and cools down to $750000$ $K$ in $\sim10$~s due to mixing with the cold clump plasma. The pressure equilibrium no longer holds and the
plasma expands in the $y-$direction leading to the distortion of the magnetic field.
Because of the inclination of the facing sides of the clumps, the magnetic field 
is distorted at two different locations with an offset along the $x-$direction.
This geometry leads to the kink of the magnetic field, as shown in Fig.~\ref{simak}b.
The magnetic field lines threading the clumps appear all similarly distorted.
After the initial kink of the magnetic field, the kink propagates along the clumps
at the kink speed, which increases once the wave leaves the clumps.
We measure the kink amplitude as the difference between the maximum and minimum $y-$coordinates
of the magnetic field line crossing the origin at $t=0$ (blue lines marking in Fig.~\ref{simak}b). 
%Fig.~\ref{simak}b shows the density and magnetic field lines at the time of
%maximum transverse displacement, which defines the kink amplitude we associate with a specific simulation.
%Throughout the simulation, the plasma temperature is affected by the blobs' collision and it follows the expected behaviour.
%The time span when the plasma temperature between the clumps remains higher than the initial temperature is $\sim35$ seconds.

\begin{figure}
\centering
\includegraphics[scale=0.35,bb=0 0 624 397]{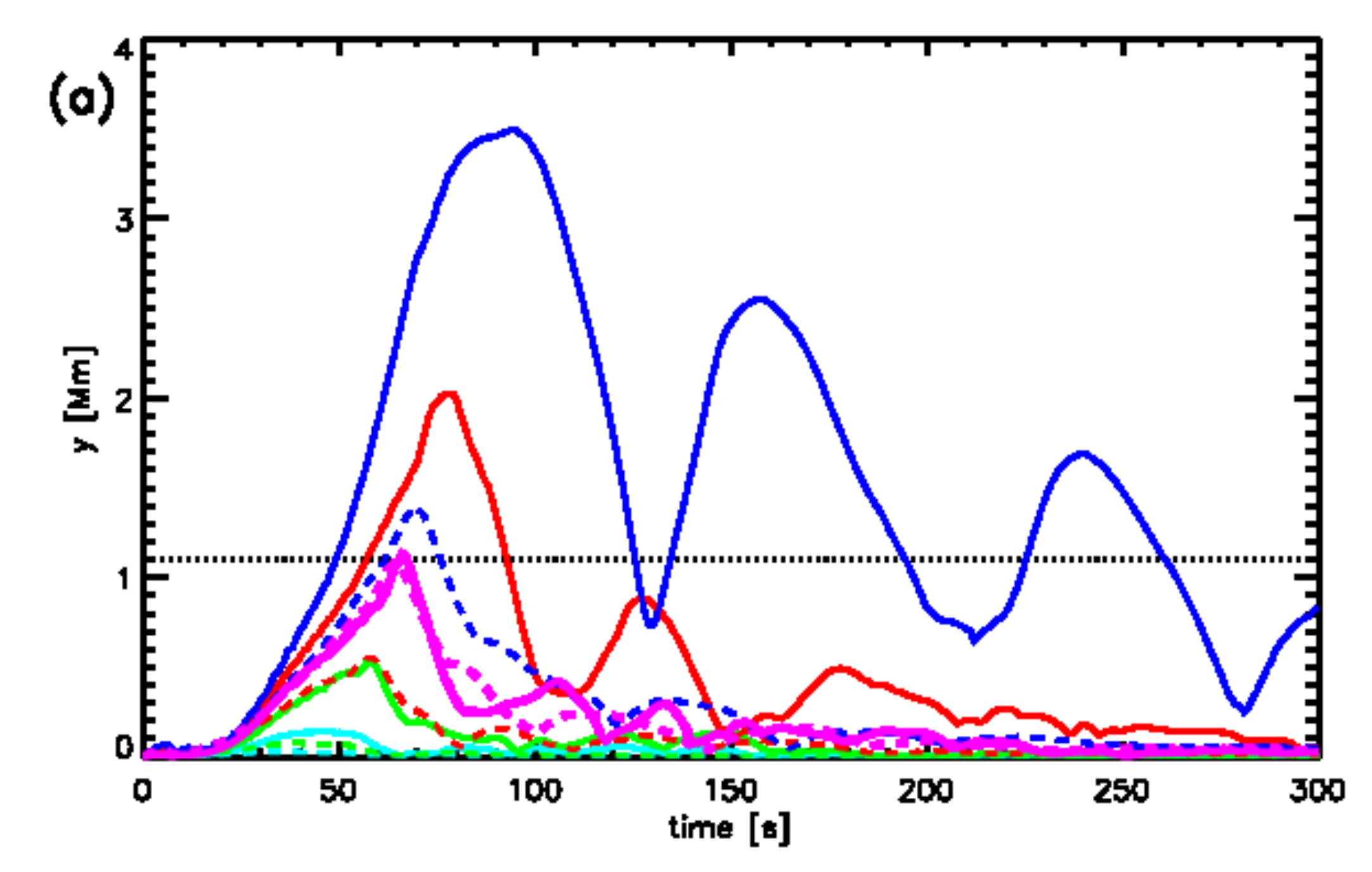}
\includegraphics[scale=0.35,bb=0 0 624 397]{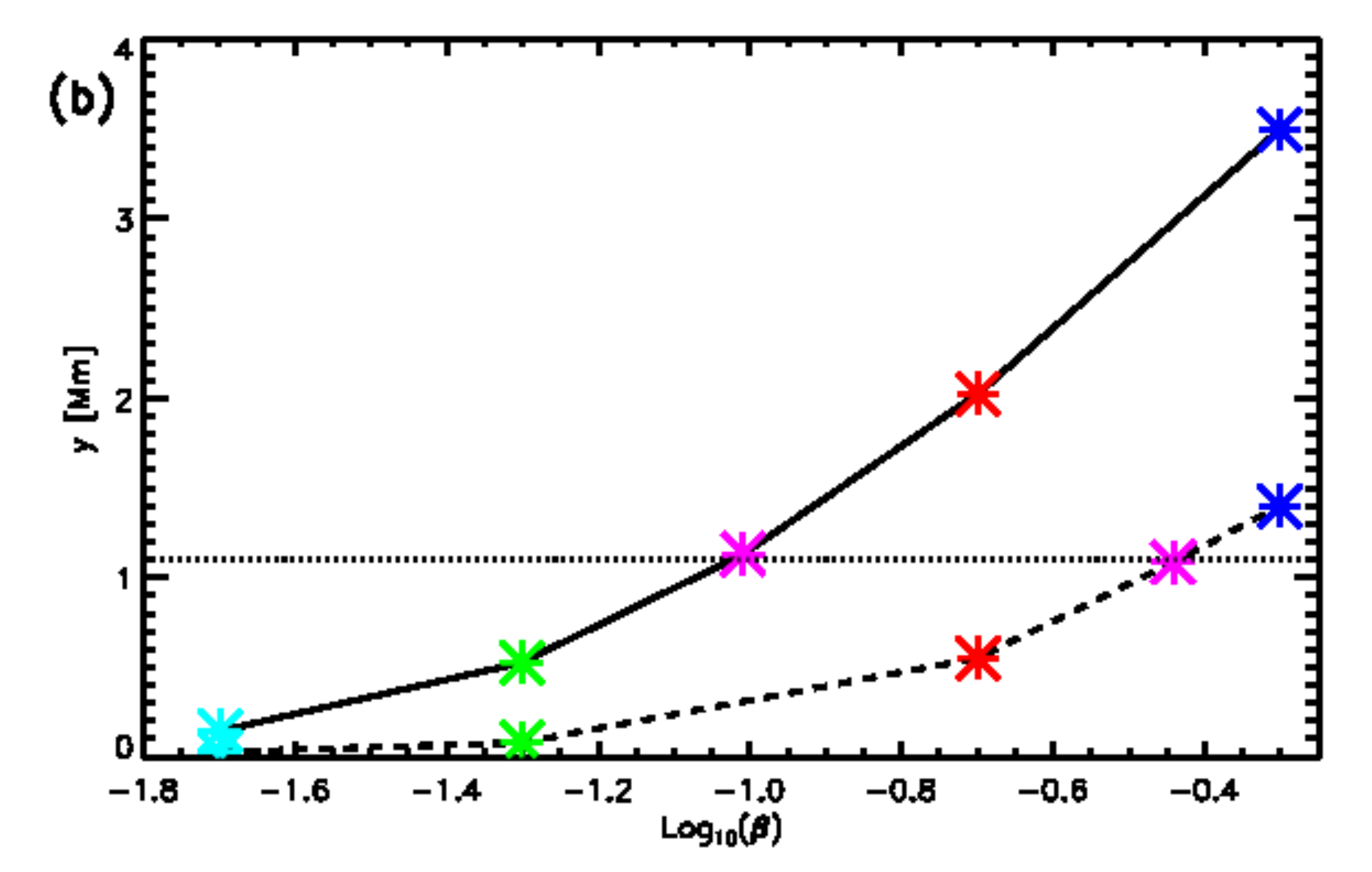}
\caption{(a) Graph of the kink amplitude as a function of time for simulations with trapezoidal clumps. Solid and dashed lines correspond, respectively, to the strong and weak case scenario. (b) Maximum kink amplitude in each simulation as function of $Log_\mathrm{10}\left(\beta\right)$. Different colours mark the $\beta$ value: light blue, green, red, and blue for $\beta=0.02, 0.05, 0.2, 0.5$, respectively. Magenta for the $\beta$ values that match best the observed kink amplitude in both scenarios.}
\label{simbeta}
\end{figure}

As the kink is generated by the imbalanced thermal pressure due to the 
compression between the clumps, the amplitude of the kink depends on the background 
plasma $\beta$.
For the strong and weak collision scenarios,
we derive the kink amplitude
as a function of $\beta$ and then, with a linear interpolation, we derive
the best $\beta$ value to match the observed kink amplitude.
%for all the simulations we have run,
%with continuous lines for the strong collision scenario and
%dashed lines for the weak collision scenario.
%Different colours mark different $\beta$ values.
We find that the kink amplitude increases until a maximum is reached and then reduces (Fig.~\ref{simbeta}a). As expected, the strong collision scenario leads to larger kink amplitudes and the higher the plasma $\beta$, the more the magnetic field is distorted. It is also evident that the wave period is longer for the larger plasma $\beta$ that implies the decrease in the Alfv\'en speed.
%The magenta lines represent the two simulations where we match the observed
%kink amplitude in the two scenarios.

Fig.~\ref{simbeta}b shows the maximum kink amplitude for the two scenarios
as a function of $\beta$.
%The horizontal dashed line marks the observed kink amplitude.
By means of this parameter space investigation, 
we derive that in the strong and weak collision scenarios
the observed kink amplitude is matched when $\beta=0.09$
and when $\beta=0.36$, respectively.
Therefore, by applying this simple model to the observed event we 
can constrain the value of the loop plasma $\beta$.
%assuming the kink oscillation is triggered in the way described here.
While the initial distortion of the magnetic field is similar to a kink mode,
%(meaning opposite $B_y$ generated at two locations at different x-coordinate),
once this travels away from the clumps, the amplitude of the remaining perturbation decreases and becomes more similar to a sausage mode (symmetric oscillation around $y=0$ axis). % with a certain $\beta$ regime.
In particular for the strong collision scenario, this occurs between $\beta=0.05$ and $\beta=0.2$. For lower $\beta$, the collision is not strong enough to produce any significant long lasting oscillation and for higher $\beta$, the post collision magnetic field becomes so entangled that it no longer behaves as a wave guide.
In the weak collision scenario we do not notice a visible persistence of sausage modes after the collision.
Similarly, as long as the clumps' collision is ongoing,
%(clumps plasma flowing towards the collision region),
the continued compression keeps the magnetic field kinked,
but the magnetic field distortion location drifts towards the external part of the clumps.
Only after the collision process is over, can the kink mode properly oscillate. Therefore, the wavelength of the initial kink oscillation depends on the length and speed of the clumps, as well as the plasma $\beta$.

To investigate the dependence on the shape of the clumps,
%which is an arbitrary parameter.
we perform two more simulations
(with the best $\beta$ values for both scenarios)
where the two clumps are symmetric and have elliptic 
facing surfaces (Fig.~\ref{simak}c). Here, the central axes
of the clumps are offset, overlapping for 75\% of their width.
In this configuration, the asymmetry that induces the kink 
is given by this offset.
The kink amplitude (Fig.~\ref{simak}d)
is found to be only $\sim10\%$ smaller
than in the simulation with trapezoidal clumps.
%Therefore the shape of the facing surfaces of the clumps 
%and their offsets are crucial to produce a kink perturbation,
%but the cause of the asymmetry is not critical.
Hence, although the presence of an asymmetry is crucial to produce a kink-like
perturbation, the exact nature of this asymmetry (shape of interface or offset) appears unimportant in the current framework. We intend to pursue a more complete parameter space investigation to identify more exactly the role and nature of this asymmetry. Future 3D simulations will address more properties of this mechanism for the generation of kink and sausage waves, including the generation of torsional waves.

% * <patrick.antolin@st-andrews.ac.uk> 2018-03-18T23:49:14.692Z:
% 
% There were incorrect values for beta before (in the caption of Fig.4). Please check that my correction is ok. 
% 
% ^.
\section{Discussion and Conclusions}\label{sec:conclude}
We have analysed coordinated observations with \textit{Hinode} and \textit{IRIS} of flows along a loop-like structure connecting a prominence with the solar surface. A collision between a downflow and an upflow is observed at estimated total speeds of $80~$km~s$^{-1}$ and $60~$km~s$^{-1}$, respectively (including Doppler and POS speeds). The densities of the flows are estimated
%through EUV absorption
to be around $6\times10^{10}-3\times10^{11}~$cm$^{-3}$. The flows are seen in SJI 2796 and 1400, indicating temperatures of $10^4-10^5$~K. At high resolution with SOT in the \ion{Ca}{2}~H line the flows appear clumpy, with widths of $0.3\arcsec-1.4\arcsec$.
%a common feature of such condensations \citep{Antolin_2015ApJ...806...81A}.
Coinciding with the time and location of collision, a bright and short-lived front is generated, indicating at least a 10 fold temperature increase. Also, at high resolution with SOT, these clumps are observed to oscillate transversely just after the collision, with an estimated total amplitude of $\approx40~$km~s$^{-1}$. We estimate a combined kinetic and enthalpy energy flux for these flows of $10^{7}-10^{8}$~erg~cm$^{-2}$~s$^{-1}$. 

Through 2D MHD numerical modelling, we have reproduced the collision between two counter-streaming flows with conditions similar to those observed. Since the clump densities are the least well-defined parameter, we allow a range of $25-100$ density contrast.
%We therefore define 2 cases, of weak and strong collision, corresponding to the low and high end values of this parameter, respectively.
Through a parameter space investigation, in order to reproduce the observed amplitude we find that the plasma $\beta$ must be confined between 0.09 and 0.36, which correspond, respectively, to magnetic field values of 6.5~G and 3.4~G.  

The modelling indicates that the presence of asymmetry between the colliding clumps leads to the {in-situ} generation of trapped and leaky MHD waves, in particular transverse and sausage, which agrees with the initially out-of-phase (radial) oscillation of strands (characteristic of the sausage mode), followed by an in-phase transverse oscillation (characteristic of the kink mode). The observed increase in period is also well explained by the modelling: the wavelength of the transverse wave is set by the length of the clumps, which increases from the time of maximum compression. Transverse MHD waves may therefore be generated in-situ in the corona through flow collision. For cool and dense prominence conditions such waves could have significant amplitudes.

The temperature at the collision can increase to coronal values, explaining the sudden intensity increase in all 3 channels. No localised signature was found in the \textit{SDO}/AIA channels (excluding AIA 304), possibly due to the increased LOS integration or long ionisation times. Nonetheless, similar signatures of counter-streaming flows and transverse MHD waves are observed at other times in this structure. The cumulative effect of such flow collisions (possibly explaining the observed background oscillation) and in-situ generated transverse MHD waves (particularly the compressive waves) may contribute to the energy balance, which may explain the EUV emission of the entire structure.

%Transverse MHD waves are present in this loop-like structure even prior to the collision. Assuming a standing fundamental mode dominates this oscillating pattern, and assuming that the core of the prominence (located at the loop apex) acts as an anchor to this loop structure (which has a length of $60-65$~Mm, taking into account the projection of the loop plane in the POS), we expect an average kink speed in the loop of $600-720~$km~s$^{-1}$. Taking an average number density along the loop of $10^{10}$~cm$^{-3}$, we can infer a magnetic field strength of $15-18$~G, which is a lower limit if we relax the assumption of the loop length. Considering the value found through the numerical modelling we can deduce that the value of $\beta$~0.024 (and therefore the strong collision case) is probably the one closest to reality for the observed case. 

\acknowledgements
This research has received funding from the UK Science and Technology Facilities Council (Consolidated Grant ST/K000950/1) and the European Union Horizon 2020 research and innovation programme (grant agreement No. 647214). VMN acknowledges the support of the  BK21 plus program through the National Research Foundation funded by the Ministry of Education of Korea. \textit{Hinode} is a Japanese mission developed and launched by ISAS/JAXA, with NAOJ as domestic partner and NASA and STFC (UK) as international partners. It is operated by these agencies in co-operation with ESA and NSC (Norway). \textit{IRIS} is a NASA small explorer mission developed and operated by LMSAL with mission operations executed at NASA Ames Research center and major contributions to downlink communications funded by ESA and the Norwegian Space Centre.
This work used the DiRAC Data Centric system at Durham University, operated by the Institute for Computational Cosmology on behalf of the STFC DiRAC HPC Facility. This equipment was funded by a BIS National E-infrastructure capital grant ST/K00042X/1, STFC capital grant ST/K00087X/1, DiRAC Operations grant ST/K003267/1 and Durham University. DiRAC is part of the National E-Infrastructure.
We acknowledge the use of the open source (gitorious.org/amrvac) MPI-AMRVAC software.%, relying on coding efforts from C. Xia, O. Porth, R. Keppens.

\end{document}